\documentclass[a4paper,11pt]{article}
\usepackage{jcappub}
\usepackage{aas_macro}

\usepackage[T1]{fontenc} 
\usepackage{graphicx}
\usepackage{orcidlink} 
\usepackage{bm}
\usepackage{epsfig}
\usepackage{amsfonts}
\usepackage{amssymb,amscd}
\usepackage{subfigure}
\usepackage{xcolor}
\usepackage{comment}
\usepackage{amsmath}
\usepackage{graphicx}
\usepackage{dcolumn}
\usepackage{hyperref}
\usepackage{xspace}
\usepackage{tabularx}
\usepackage{booktabs}
\usepackage{multirow}
\usepackage{fancyheadings}
\usepackage{fancyhdr}
\usepackage{soul}
\usepackage{theorem}
\usepackage{moreverb}
\usepackage{euscript}
\usepackage{psfrag}
\usepackage{slashed}
\usepackage{mathtools}
\usepackage{makecell}
\usepackage[flushleft]{threeparttable}
\usepackage{adjustbox}
\usepackage{lineno}

\begin{document}
\newcommand{\EM}[1]{\textcolor{teal}{[EM: #1]}}
\newcommand{\DM}[1]{\textcolor{purple}{[DM: #1]}}
\newcommand{\CG}[1]{\textcolor{orange}{[CG: #1]}}

\def\lat{\textit{Fermi-LAT}\xspace}
\def\hess{H.E.S.S.\xspace}
\newcommand{\green}[1]{{\color{green}{#1}}}
\newcommand{\blue}[1]{{\color{blue}{#1}}}

\title{Sensitivity to Axion-like Particle dark matter with very-high-energy gamma-ray observations of Active Galactic Nuclei located behind Galaxy Clusters}

\author[a]{Cervane Grimaud$^{\orcidlink{0000-0001-7736-7730}}$}

 \author[b]{Denys Malyshev$^{\orcidlink{0000-0001-9689-2194}}$}

 \author[a]{Emmanuel Moulin$^{\orcidlink{0000-0003-4007-0145}}$}

\affiliation[a]{Irfu, CEA Saclay, Universit\'e Paris-Saclay, F-91191 Gif-sur-Yvette, France}

\affiliation[b]{Institut f\"ur Astronomie und Astrophysik, Universit\"at T\"ubingen, Sand 1, D 72076 T\"ubingen, Germany}

\emailAdd{cervane.grimaud@cea.fr}
\emailAdd{denys.malyshev@astro.uni-tuebingen.de}
\emailAdd{emmanuel.moulin@cea.fr}

\date{\today}

 \abstract{
 Axion-Like-Particles (ALPs) are hypothetical pseudo-scalar particles actively searched as light dark matter candidates. The coupling of ALPs to photons can give rise to distinctive spectral features in the observed gamma-ray spectrum of astrophysical sources.
 We perform a forecast study on the sensitivity to ALP-photon interactions using stacked mock observations of selected active galactic nuclei (AGNs) located behind galaxy clusters (GC). The ALP-photon conversion in the magnetic fields of galaxy clusters give rise to absorption-like features in AGN spectra that are subject to large variance in their prediction for individual sources. We consider here a stacking analysis of multiple AGN-cluster pairs, which yields a more controlled prediction of the expected ALP-induced spectral patterns in the observed gamma-ray spectra.
 Using realistic mock observations of selected \lat AGNs  by ongoing Imaging Atmospheric Cherenkov Telescopes such as H.E.S.S., MAGIC and VERITAS, we provide a careful assessment of the expected sensitivity of a combined statistical analysis of many AGN-GC pairs, together with the impact of modelling and instrumental uncertainties. 
The sensitivity reaches ALP-photon couplings down to 6$\times$10$^{-13}$ GeV$^{-1}$ for an ALP mass of 3$\times$10$^{-8}$ eV, and is currently statistically dominated indicating further improvements from more observations.
Such a stacking analysis approach enables exploration of the yet-uncharted ALP dark matter parameter space in the 10$^{-8}$ - 10$^{-7}$ eV mass range.
 }

\maketitle

\section{Introduction}
\label{sec:intro}
Among the compelling dark matter candidates are axion-like particles (ALPs) which arise in several Beyond-the-Standard-Model theories~\cite{Arias:2012az,PeterSvrcek_2006,PhysRevD.81.123530} as a pseudo–Nambu-Goldstone boson resulting from the breaking of a U(1) symmetry. Provided ALPs are non-thermally produced via mechanisms including misalignment, thermal inflationary production, and from the decays of heavier particles (see, \textit{e.g.}, Ref.~\cite{Sikivie:2006ni}), they can constitute the observed relic abundance of dark matter in the Universe. As opposed to axions intended to solve the strong CP problem~\cite{pq, kim, SHIFMAN1980493, DINE1981199, Zhitnitsky:1980tq}, the ALP coupling is not correlated anymore to mass, which broadens significantly the available parameter space. 
A generic feature of ALPs is their coupling to a two-photon vertex, which in the presence of an external magnetic field can lead to ALP-photon oscillations~\cite{Raffelt:1987im}, which has been extensively used to probe ALP properties with GeV-TeV gamma rays, see, \textit{e.g.}, Refs.~\cite{Hooper:2007bq,HESS:2013udx,Fermi-LAT:2016nkz,MAGIC:2024arq,VERITAS:2025ett}.

The existence of ALPs has strong implications for different astrophysical phenomena, such as the increased transparency of the Universe to high energy photons~\cite{DeAngelis:2007dqd,Mirizzi:2007hr,Simet:2007sa} and on the evolution of stars~\cite{Ayala:2014pea,Giannotti:2015kwo,MillerBertolami:2014rka}. Dedicated techniques deployed to look for ALPs include light-shining-through-wall experiments, birefringence and dichroism in light polarisation, resonant cavities, axion helioscopes and haloscopes, see, \textit{e.g.}, thorough reviews in Refs.~\cite{Irastorza:2018dyq,Sikivie:2020zpn}.

The probability of the ALP production via photon-ALP conversion in an external magnetic field is maximized for large values of the magnetic field and long propagation baselines. Astrophysical environments opportunely offer bright sources of photons, a wide range of magnetic fields and very long baselines. Active Galactic Nuclei (AGN) such as blazars are among the best sources of very-high-energy (VHE, E $\gtrsim$ 100 GeV) gamma rays to probe ALP properties. During its propagation from a blazar to the Earth, a VHE gamma ray traverse regions with different magnetic fields such as the ones at the source itself, in the relativistic jet, possibly the one of the galaxy cluster in which the AGN is residing or being along the line of sight, the intergalactic magnetic field (IGMF), and the Milky Way one. Energy dependent photon-ALP conversion can occur in each of the described regions resulting in a feature(s) in otherwise smooth VHE AGN spectrum. The characteristic energy of the feature depends on the strength and spatial scale of the magnetic field, which allows to expect that the discussed fields affect different energy bands and not result in superposition of effects at the same energy. The accuracy of the knowledge of the discussed magnetic fields varies however significantly from a region to another which makes certain regions more preferable for studying photon-ALP conversion. 

The IGMF strength is difficult to measure directly and only upper limits of the order of 1~nG are obtained from Faraday rotation measurements~\cite{Pshirkov:2015tua}. Early studies have considered conversion in the IGMF assuming this 1~nG upper limit, however this poorly known value impacts significantly the robustness of the photon-ALP conversion expectations in such a field. 
IGMF present in voids would deflect the trajectories of electrons and positrons produced as secondary emission in the gamma-ray propagation, such that the secondary emission in gamma rays is not directed along the primary gamma-ray beam anymore. Observing this suppression in the direction of the primary gamma-ray beam has been used to establish the existence of non-zero magnetic field in the voids~\cite{Neronov:2010gir,Taylor:2011bn} leading to a recent lower bound derived at 10$^{-17}$ G~\cite{Blunier:2025ddu}. 

An alternative approach is to search for conversion in relativistic jets of AGN  where magnetic fields of the order of 1~G can be expected on distance of $\sim$100~pc. For this method, the computation of the photon survival probability requires an accurate determination of the gamma-ray path in the jet itself through the modelling of the magnetic field in the blazar’s jet to compute the survival probability, see, for instance,  Ref.~\cite{Davies:2020uxn}.

Galaxy clusters (GC) offer environments where the strength of the magnetic field can be derived from the observed synchrotron emission, the inverse Compton emission, or, more experimentally, with the Faraday RM effect, see, \textit{e.g.} Ref.~\cite{Govoni:2004as}.
In particular, the latter approach was used to derive magnetic field strength of the order of $\mu$G in the Perseus~\cite{2006MNRAS.368.1500T} and Coma~\cite{Bonafede:2010xg} clusters.
Looking for photon-ALP conversion in the GC magnetic field offers the advantage of the existing estimation of the magnetic field strength, but lacks, however, the reliable measurement of the small-scale properties of the field such as orientation of the field in domains and measurement of domains lengths. 

Since photon-ALP conversion is strongly dependent on these properties, reliable estimation of the energy-dependence of a photon-ALP conversion probability is almost impossible in a single AGN-cluster pair.
For a single pair the probability of the photon-ALP conversion in the realistic magnetic field of the cluster is an oscillating (close to characteristic energy) function, with the exact shape/positions of oscillations determined by the small-scale properties of the field. Derivation of limits on ALP parameters from (non-)observation of such oscillations in spectrum of an single AGN behind a cluster of galaxies requires thus a complex procedure of marginalisation over a large number of (ideally -- all possible) small-scale configurations of the magnetic field in a cluster~\cite[see e.g.][]{Fermi-LAT:2016nkz,2018arXiv180504388M,2025PhRvD.112g5008G}, due to which a relatively weak bounds are derived.

To overcome this limitation, stacking analyses of multiple AGN-galaxy cluster pairs have been proposed~\cite{Malyshev:2025iis}.
Averaging over many independent AGN-GC pairs results in a smooth, predictable spectral imprint. This stacking technique significantly enhances the statistical power and enables the derivation of robust constraints on ALP parameters. 
We will show here that a stacked log-likelihood ratio test statistic analysis of mock observational datasets by ongoing Imaging Atmospheric Cherenkov Telescopes (IACTs) enable to probe the ALP-photon couplings expected for dark matter ALP in the yet uncharted 10-100 neV mass range. The key findings shown by our analysis are: \textit{(i)} Observations of AGN located behind galaxy clusters with currently ongoing IACTs possess the required sensitivity to probed yet-uncharted of ALP dark matter in the 10-100 neV mass range. \textit{(ii)} The sensitivity reach is statistics dominated, implying that increased data collection towards selected AGNs with ongoing IACTs remains important. \textit{(iii)} The impact of the extragalactic background light (EBL) model on the sensitivity reach is limited provided statistically homogeneous datasets for AGNs distributed over a range of redshifts. 

The paper is organized as follows. 
Section~\ref{sec:selection} describes the selection procedure of AGNs from \lat catalog and their relevance for H.E.S.S., MAGIC and VERITAS observations. In section~\ref{sec:models}, we outline the key ingredients to derive the expected pattern for the survival probability of VHE gamma rays for the observations of stacked AGN-GC pairs with IACTs together with the expected VHE fluxes. 
Section~\ref{sec:analysis} presents the statistical analysis framework devised to combine mock observations of AGN-GC pairs
and compute the expected sensitivity to the ALP-photon coupling.
Section~\ref{sec:results} is devoted to the results 
together with a determination of the systematic uncertainties associated with the modelling.  Finally, Section~\ref{sec:conclusions} is devoted to the conclusions.

\section{Selection of AGN-GC pairs for mock observations}
\label{sec:selection}

\subsection{Sample of Fermi-LAT AGNs from 4FGL and galaxy cluster catalogs}
\label{sec:flatagn}
In order to identify the AGNs located behind clusters of galaxies we cross-correlate the high-latitude Fourth Catalog of Active Galactic Nuclei (Galactic latitudes $|b|>10^\circ$) \texttt{4LAC-DR3-h}~\citep{4lacdr3} of gamma-ray bright AGNs detected by \lat with the catalogues of Sunaev-Zeldovich, optically and X-ray identified clusters of galaxies~\citep{2018MNRAS.475..343W,planck_2015sz,erass1_clusters}. Similarly to Ref.~\cite{Malyshev:2025iis}, among AGNs and clusters with known redshifts $z$, we select pairs for which \textit{(i)}: $z_{\rm AGN}\geq z_{\rm cluster}$; \textit{(ii)}: the line of sight from AGNs passes by not more than 500~kpc comoving distance from the cluster's center. We note, that selection of 500~kpc is a conservative selection of the distance to the cluster's center, as the magnetic field in clusters of galaxies reportedly continues to larger distances up to a few Mpc~\citep{1989Natur.341..720K, 2015JCAP...01..011K}. In addition to the selected pairs identified in such a way pairs, we included into the considered sample NGC~1275 and M~87 located close to the centers of nearby Perseus and Virgo clusters, respectively. 

Table~\ref{tab:agn_sample} provides the \lat names of the selected AGNs, their coordinates, redshifts,   best-fit spectral parameters, as well as their multiwavelength counterparts if any.
Such a selection provides a sample of 29 AGNs for which the best-fit spectral parametrization is a pure power law.
In what follows, to conservatively estimate the number of sources detectable in the TeV band we further restrict our sample to AGNs present in the 3FHL catalog~\citep{Fermi-LAT:2017sxy} of sources detected by \lat above 10~GeV. Such a conservative approach, resulted in 16 AGNs promising for TeV observations, see Table~\ref{tab:agn_visibility}.
 
For these AGNs we explicitly adopted the reported in 3FHL spectral index $\Gamma$ and flux $F$ at 100~GeV. For completeness, for objects not present in 3FHL but having best-fit power-law spectral model in 4LAC catalogue~\citep{4lacdr3} in Table~\ref{tab:agn_sample} we present the best-fit model estimated spectral parameters --
the flux at 100~GeV and corresponding uncertainty. We marked the flux estimated in such a way with a dagger $^\dagger$ in Table~\ref{tab:agn_sample}. We additionally marked with $^{\dagger\dagger}$ sources with non-zero analysis flag in 4LAC catalogue which indicates possible issues with the reported in catalogue values.
\begin{table}[ht!]
\footnotesize
    \centering
    \begin{tabular}{l|c|c|c|l|l}
    \hline\hline
    \lat name & RA & Dec. & z & Spectral parameters & Other names \\ \hline\hline
4FGL J0014.2+0854 & 3.57 & 8.91 & 0.16 &  $\Gamma = 2.5$ , F=0.8 $\pm$ 0.2$^{\dagger\dagger}$ &  \\
4FGL J0038.2-2459 & 9.56 & -24.99 & 1.20 & LogParabola in 4LAC &  PKS 0035-252\\
4FGL J0049.0+2252 & 12.25 & 22.87 & 0.26 &  $\Gamma = 2.3$ , F=4.0 $\pm$ 0.9$^\dagger$ &  \\
4FGL J0132.7-0804 & 23.18 & -8.07 & 0.15 &  $\Gamma = 2.0$ , F=27.2 $\pm$ 4.0$^\dagger$ &  \\
4FGL J0317.8-4414 & 49.47 & -44.24 & 0.08 &  $\Gamma = 2.0$ , F=12.7 $\pm$ 3.1$^\dagger$ &  \\
4FGL J0617.7-1715 & 94.43 & -17.25 & 0.10 &  $\Gamma = 2.9$ , F=3.8 $\pm$ 2.7 & 3FHL J0617.6-1715 \\
4FGL J0912.5+1556 & 138.14 & 15.93 & 0.21 &  $\Gamma = 1.8$ , F<7.3 ($\sim 5.2$) & 3FHL J0912.4+1555 \\
4FGL J0914.4+0249 & 138.61 & 2.83 & 0.43 &  $\Gamma = 2.6$ , F=0.4 $\pm$ 0.1$^\dagger$ &  \\
4FGL J1010.8-0158 & 152.70 & -1.98 & 0.90 &  $\Gamma = 2.5$ , F=1.2 $\pm$ 0.1$^\dagger$ &  \\
4FGL J1013.7+3444 & 153.45 & 34.74 & 0.21 &  $\Gamma = 2.5$ , F=2.3 $\pm$ 0.2$^\dagger$ &  \\
4FGL J1058.4+0133 & 164.62 & 1.56 & 0.89 &  $\Gamma = 2.9$ , F=6.5 $\pm$ 4.2 & 3FHL J1058.4+0133 \\
4FGL J1202.5+3852 & 180.65 & 38.88 & 0.81 &  $\Gamma = 2.3$ , F=2.9 $\pm$ 0.4$^{\dagger\dagger}$ &  \\
4FGL J1213.0+5129 & 183.26 & 51.48 & 0.80 &  $\Gamma = 2.0$ , F<6.7 ($\sim 3.9$) & 3FHL J1213.0+5128 \\
4FGL J1303.0+2434 & 195.76 & 24.58 & 0.99 &  $\Gamma = 2.9$ , F<5.4 ($\sim 2.9$) & 3FHL J1303.0+2435 \\
4FGL J1353.2+3740 & 208.30 & 37.68 & 0.22 & LogParabola in 4LAC &  \\
4FGL J1508.8+2708 & 227.20 & 27.14 & 0.27 &  $\Gamma = 2.0$ , F=3.2 $\pm$ 2.8 & 3FHL J1508.7+2708 \\
4FGL J1516.8+2918 & 229.21 & 29.31 & 0.13 &  $\Gamma = 2.2$ , F=3.5 $\pm$ 0.6$^\dagger$ &  \\
4FGL J1615.6+4712 & 243.92 & 47.20 & 0.20 &  $\Gamma = 3.0$ , F<6.1 ($\sim 0.8$) & 3FHL J1615.4+4711 \\
4FGL J2041.9-3735 & 310.48 & -37.59 & 0.10 &  $\Gamma = 1.6$ , F=11.0 $\pm$ 5.9 & 3FHL J2041.9-3734 \\
4FGL J2314.0+1445 & 348.51 & 14.75 & 0.16 &  $\Gamma = 2.0$ , F=16.6 $\pm$ 7.1 & 3FHL J2314.0+1445 \\
4FGL J2321.9+2734 & 350.49 & 27.57 & 1.25 &  $\Gamma = 3.8$ , F<5.2 ($\sim 0.9$) & 3FHL J2321.9+2735 \\
4FGL J2336.6+2356 & 354.15 & 23.94 & 0.13 &  $\Gamma = 2.0$ , F=30.9 $\pm$ 4.0$^\dagger$ &  \\
4FGL J2338.9+2124 & 354.74 & 21.40 & 0.29 &  $\Gamma = 2.3$ , F=10.6 $\pm$ 5.7 & 3FHL J2338.9+2123 \\
4FGL J0303.3-7913 & 45.84 & -79.22 & 1.12 &  $\Gamma = 2.5$ , F=1.1 $\pm$ 0.1$^\dagger$ &  \\
4FGL J0309.4-4000 & 47.36 & -40.01 & 0.19 &  $\Gamma = 1.9$ , F=20.6 $\pm$ 4.4$^\dagger$ &  \\
4FGL J0654.4+4514 & 103.61 & 45.24 & 0.93 &  $\Gamma = 2.5$ , F=3.5 $\pm$ 2.5 & 3FHL J0654.4+4514 \\
4FGL J1242.9+7315 & 190.75 & 73.26 & 0.08 &  $\Gamma = 1.6$ , F<5.6 ($\sim 3.9$) & 3FHL J1243.0+7316 \\
4FGL J0319.8+4130 & 49.96 & 41.51 & 0.02 &  $\Gamma = 2.9$ , F=49.1 $\pm$ 11.8 & NGC 1275, 3FHL J0319.8+4130 \\
4FGL J1230.8+1223 & 187.71 & 12.39 & 0.00 &  $\Gamma = 1.8$ , F=19.3 $\pm$ 8.0 & M87, 3FHL J1230.8+1223 \\
4FGL J1144.9+1937 & 176.24 & 19.63 & 0.02 & $\Gamma = 1.7$ , F=12.7 $\pm$ 6.0 & 3FHL J1145.0+1935 \\
4FGL J0805.2-0110 & 121.30 & -1.18 & 1.39 & LogParabola in 4LAC &  \\
4FGL J1116.6+2915 & 169.16 & 29.26 & 0.05 &  $\Gamma = 1.6$ , F=74.3 $\pm$ 29.1$^{\dagger\dagger}$ &  \\
\hline\hline
    \end{tabular}
    \caption{Sample of selected AGNs from the \texttt{4LAC-DR3-h} catalogue. The table summarizes 4FGL-catalogue name of the source in the first column, together  with  its RA-Dec coordinates in the second and third columns, respectively. The fourth column provides the redshif $z$ of the source. The spectral parameters, spectral index $\Gamma$ and flux normalisation $F$ at 100 GeV in units of $10^{-13}$~erg/cm$^2$/s, are given in the fifth column.
    In the last column are summarized, if any, conventional name of the source, its 3FHL name and spectral index $\Gamma$ at $>10$~GeV as reported in \texttt{3FHL} catalogue~\cite{Fermi-LAT:2017sxy}.  
    The estimated flux was calculated based on the power-law fit to the 3FHL data points.  
    Two relevant sources (M~87 and NGC~1275) are already detected. 
    $^\dagger$ -- the source is present in 4LAC catalogue with a power-law best-fit model and not present in 3FHL. $^{\dagger\dagger}$ -- non-zero analysis flag in 4LAC, indicating possible issues with the analysis.}
    \label{tab:agn_sample}
\end{table}

\begin{table}[h!]
\footnotesize
    \centering
    \begin{center}   
    \begin{tabular}{l|c|ccc}
    \hline\hline
\lat name & Visibility & \multicolumn{3}{c}{Number of hours available per year for ZA$<$60$^\circ$} \\
\cline{3-5}
\hline
& & H.E.S.S. & MAGIC & VERITAS \\
\hline
4FGL J0617.7-1715 & H + M + V &579.0&399.05&260.1 \\
4FGL J0912.5+1556 & H + M + V &416.3&587.2&300.25\\
4FGL J1058.4+0133 & H + M + V &506.4&517.6&314.8\\
4FGL J1213.0+5129 & M + V &&593.0&269.2\\
4FGL J1303.0+2434 & H + M + V &339.6&535.9&267.3\\
4FGL J1508.8+2708 & H + M + V &313.25&513.8&277.6\\
4FGL J1615.6+4712 & M + V &&555.2&292.4\\
4FGL J2041.9-3735 & H &672.9&&\\
4FGL J2314.0+1445 & H + M + V &469.05&596.8&363.05\\
4FGL J2321.9+2734 & H + M + V &329.6&637.0&401.35\\
4FGL J2338.9+2124 & H + M + V &410.1&626.6&404.95\\
4FGL J0654.4+4514 & M + V &&681.25&331.2\\
4FGL J1242.9+7315 & M + V &&581.8&298.3\\
4FGL J0319.8+4130 & M + V &&702.95&372.3\\
4FGL J1230.8+1223 & H + M + V &450.3&535.3&308.9\\
4FGL J1144.9+1937 & H + M + V &389.8&567.2&306.15\\
 \hline\hline
\end{tabular}
\end{center}
    \caption{Visibility of the sample of 16 selected AGNs. The first and second columns give the  name of the source from the \texttt{4LAC} catalogue, and the potential visibility by the ongoing IACTs H.E.S.S. (H), MAGIC (M) and VERITAS (V), respectively. The last column indicates the number of hours of observations under darkness available per year for each considered IACT with an observational zenith angle \textbf{lower than 60$^\circ$} tcomputed from Ref.~\cite{visibility}.}
    \label{tab:agn_visibility}
\end{table}

\subsection{Selection for IACT observations}
\label{sec:obs}
Aiming to study the sensitivity of IACTs to the detection of ALPs we additionally restrict the selected sample to the objects visible by currently operating IACTs.
In our analysis, we will therefore focus on mock observations by H.E.S.S., MAGIC and VERITAS above 100 GeV.

H.E.S.S. is an array of five IACTs, four 12m-diameter and one 28m-diameter, located in the Khomas Highlands of Namibia, 23$^\circ$16$'$18$''$S, 16$^\circ$30$'$0$''$E, at 1.8 km a.s.l.~\cite{hess} with more than 1300 hours of data taking per year. 
The energy-dependent instrument response functions (IRFs) are extracted from Ref.~\cite{HESS:2022ygk} as representative of H.E.S.S. observations at zenith angle of 20$^\circ$. We note that these H.E.S.S. IRFs are obtained from averaging over several observations taken at different zenith angles and for different instrumental and atmospheric conditions.
MAGIC is composed of two 17m-diameter telescopes situated 2.2 km a.s.l, 28$^\circ$45$'$43$''$N, 17$^\circ$53$'$24$''$W , at the Roque de los Muchachos Observatory, on the Canary Island of La Palma, Spain~\cite{magic}.
The energy-dependent IRFs are extracted from Ref.~\cite{Abe:2023kvd} as representative of MAGIC observations at zenith angle of 20$^\circ$.
VERITAS comprises an array of four IACTs located in Southern Arizona at 31$^\circ$40$'$30$''$N, 110$^\circ$57$'$07$''$W, 1.3 km a.s.l.~\cite{veritas}.
The energy-dependent IRFs are extracted from Ref.~\cite{Adams:2021hzq} a representative of VERITAS observations under zenith angle of 20$^\circ$. According to Ref.~\cite{Neronov:2025smk},
$\gtrsim$100 AGNs have been detected in VHE gamma rays so far.
We select here AGN included in the 3FHL catalog with a visibility at more than $30^\circ$ above horizon for at least one of the considered IACTs. Namely, for each observatory, we select objects with the declination $Dec$, for which the minimal zenith angle $MZA \equiv |-24.6-Dec|$ is $MZA<60^\circ$. This results in 11 AGN-cluster pairs suitable for further analyses with \hess and 15 pairs both suitable for MAGIC and VERITAS. 
Mock observations by a given IACT assume 50 hours of observations of the selected AGN given in Table~\ref{tab:agn_sample}.
Such an observation time is realistic given the yearly selected-AGN visibility by IACTs and the observational time that can be dedicated to these objects due to prioritization of observations in the relevant right-ascension bands.

Some AGNs can be observed by more than one observatory, and even by the three observatories, which leads to 41 independent observational mock datasets.
Table~\ref{tab:agn_visibility} gives the visibility of the selected sources
by H.E.S.S., MAGIC and VERITAS together with the number of hours of observations under darkness condition available per year for each IACT for zenith angle lower than 30$^\circ$.

\subsection{Modelling the galaxy cluster magnetic field}
\label{sec:b_field}
The presence of magnetic fields in some GCs is supported by observations of regions emitting diffuse synchrotron emission, thus revealing the presence of relativistic electrons and $\sim \mu$G magnetic fields. These regions are classified into radio halos, relics and mini-halos, depending on their size and position within the GC. Estimates of the strength of the magnetic field from these three types of targets can be reconciled through magnetohydrodynamic simulations, all within the range of 0.1 - 10 $\mu$G~\cite{Govoni:2004as}.

We consider in the following, cool core GCs which show a good agreement with the assumption of the magnetic field profile to be dependent on the electron density, and estimated to $\sim$10 $\mu$G in the central regions. The intracluster environment makes use of the Coma cluster (Abell 1656) as a benchmark~\cite{Colafrancesco:2005ji,Bonafede:2010xg} where a central magnetic field value of 5.2$\,\mu$G is taken~\cite{Bonafede:2010xg}. Based on Faraday rotation measurements,  
central values as large as 25$\,\mu$G are found in the cool-core Perseus cluster~\cite{2006MNRAS.368.1500T}. 
The well-characterized X-ray and radio features of the Coma cluster, along with extensive Faraday Rotation Measure data, make it ideal for simulating photon-ALP mixing. The thermal electron density is well represented by a standard $\beta$-model:
\begin{equation}
n_e(r) = n_0 \left(1 + \frac{r^2}{r_c^2} \right)^{-3\beta/2},
\end{equation}
with $n_0 = 3.44 \times 10^{-3} \, \text{cm}^{-3}$, $r_c = 291 \, \text{kpc}$, and $\beta = 0.75$~\cite{Bonafede:2010xg,2003MNRAS.343..401L}. The magnetic field strength scales with the local electron density as:
\begin{equation}
B(r) = B_0 \left( \frac{n_e(r)}{n_0} \right)^{0.67},
\end{equation}
and vanishes beyond a radial distance corresponding to half the total simulated path length, effectively confining the magnetic field to the central, magnetized core of the cluster. In absence of a large number of reliable measurements of magnetic fields in other galaxy clusters~\cite{Malyshev:2025iis} we consider this profile as reference one and characteristic for majority of clusters. Such a choice allows also direct comparison of the derived IACTs sensitivity estimation to the limits from lower-energy \lat observations~\cite{Malyshev:2025iis}.

\section{Expected VHE gamma-ray signals}
\label{sec:models}

\subsection{Photon survival probability}
Following Refs.~\cite{Raffelt:1987im,Sikivie:2020zpn},  photons travelling in an external magnetic field \textbf{B}, in the z direction, can oscillate into an ALP via the equation of motion:
\begin{equation}
\Big( E - i\partial_z - \mathcal M(m_a, g_{a\gamma\gamma}, \bm{B}_\perp(z)) \Big)\left(\begin{array}{l}A_x \\ A_y \\ \,a \end{array}\right) = 0, 
\label{eq:eom}
\end{equation}
where $A_x$ and $A_y$ correspond to the two linear photon polarization states, perpendicular to the propagation direction $z$, and $E$ is the photon or axion energy. 
Note that the component of B parallel to the direction of propagation does not induce photon-ALP mixing. 
The mixing matrix $\mathcal M(m_a, g_{a\gamma\gamma}, \bm{B}_\perp(z))$ depends on the strength and orientation of the magnetic field in the transversal plane, on the ALP mass $m_a$ and the coupling constant $g_{a\gamma\gamma}$. For GeV-TeV gamma-ray energies of interest here, the Cotton-Mouton and the Faraday terms can be neglected. The Faraday rotation term is relevant when analyzing polarized sources of photons but it plays no role in the case studied here. Therefore, the matrix is given by :
\begin{equation}
\label{eq:mixing_matrix}
    \mathcal M = \left(\begin{array}{ccc}
    \Delta_{pl} & 0 & \Delta_{a\gamma}\cos\phi \\
    0 & \Delta_{pl} & \Delta_{a\gamma}\sin\phi \\
\Delta_{a\gamma}\cos\phi & \Delta_{a\gamma}\sin\phi & \Delta_a 
\end{array}\right),
\end{equation}
with $\cos\phi={\bm B}_{\perp}\cdot{\bf e}_x/B_{\perp}
= \sqrt{1-\sin^2\phi}$.
Under these assumptions, the orthogonal component of the axion potential decouples and the equations of motion simplify as the 2 × 2 mixing problem.

By analogy to the neutrino mixing states, the probability of the photon-ALP conversion in an homogeneous magnetic field for a travelled distance $s$ is given by~\cite{Raffelt:1987im}:
\begin{equation}
P_{\gamma a} = \frac{1}{2} \rm sin^2(2\theta) \rm sin^2\Bigg( \frac{2\pi s}{\lambda_{\rm osc}}\Bigg) = (\Delta_{a\gamma} s)^2 \frac{sin^2(\Delta_{\rm osc} s/2)}{(\Delta_{\rm osc} s/2)^2}\,\,\, {\rm with\,\,\,} \theta = \frac{1}{2}\rm arctan\Bigg(\frac{2\,B\,g_{a\gamma\gamma}}{m_a^2-\omega_{pl}^2}\Bigg)\, .
    \label{eq:oscp}
\end{equation}
Here $\Delta_{\rm osc}$ is the oscillation wave number given by 
$\Delta_{\rm osc}^2 = (\Delta_{\rm a}-\Delta_{\rm pl})^2 + 4
\Delta_{\rm a\gamma}^2$, where $\Delta_{a\gamma} = 1/2 g_{a\gamma\gamma} B_{\rm T}$, $\Delta_{a} = -m_{\rm a}^2/2E$, and $\Delta_{\rm pl} = -\omega^2_{\rm pl}/2\omega$.
Here we consider gamma rays with energies from 100 GeV up to 100 TeV, such that the  plasma oscillation number is negligible. The oscillation wave number therefore simplifies to :
\begin{equation}
\Delta_{\rm osc} \simeq \sqrt{\Delta_{\rm a}^2 - 4\Delta_{\rm a\gamma}^2} = 2 \Delta_{\rm a\gamma} \sqrt{1-(E/E_{\rm c})^2} \quad \rm with \quad E_{\rm c} \equiv \frac{\Delta_{\rm {a}}}{2\Delta_{a\gamma}}E \, ,
\label{eq:deltaosc}
\end{equation}
where the critical energy $E_c$ represents the energy where the conversion probability becomes sizeable. 
Neglecting the plasma frequency that does not impact substantially in the energy range of interest, the critical energy can be rewritten:
\begin{equation}
    E_{\rm c} \simeq 0.25\ \rm TeV \Bigg(\frac{m_a}{10^{-8}\ eV}\Bigg)^2 \Bigg(\frac{3\times10^{-12}\ \rm GeV}{g_{a\gamma\gamma}}\Bigg) \Bigg(\frac{3\ \mu G}{B_{\rm T}}\Bigg)
\label{eq:ecrit}
\end{equation}

In astrophysical environments, the magnetic field is turbulent, \textit{i.e.}, the field orientation changes along the photon propagation path. The propagation over many B-field domains needs to account for possibly different domain sizes, non-trivial dependency of the magnetic field on the domain position in the cluster as well as on the orientation of the magnetic field in each domain. 

\begin{figure}
        \centering
        \includegraphics[keepaspectratio, width=0.495\linewidth, clip]{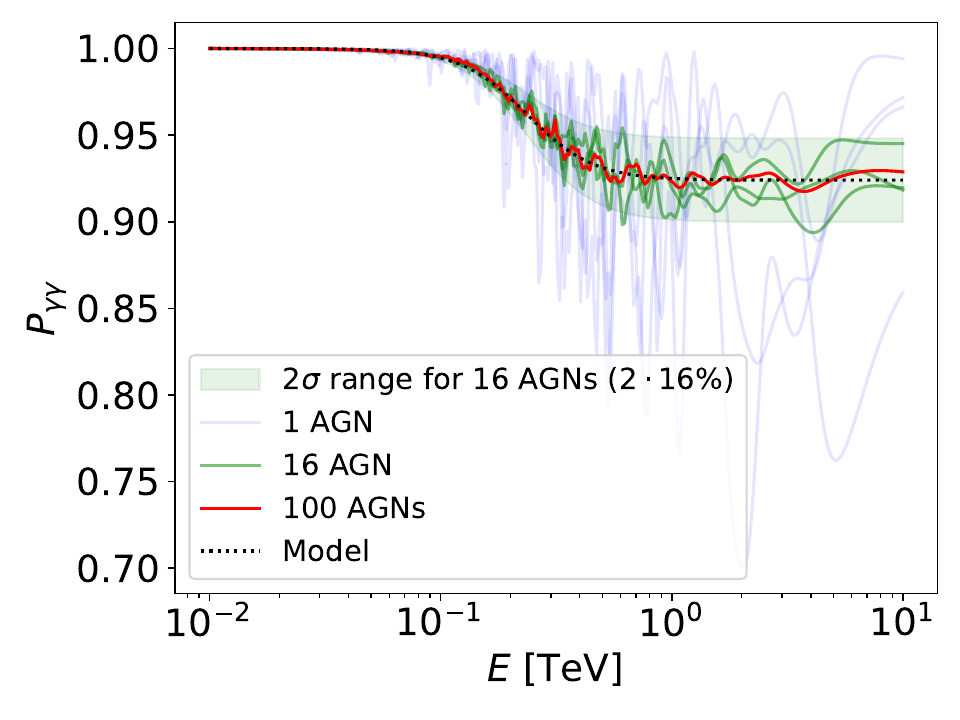}
        \includegraphics[keepaspectratio, width=0.495\linewidth, clip]{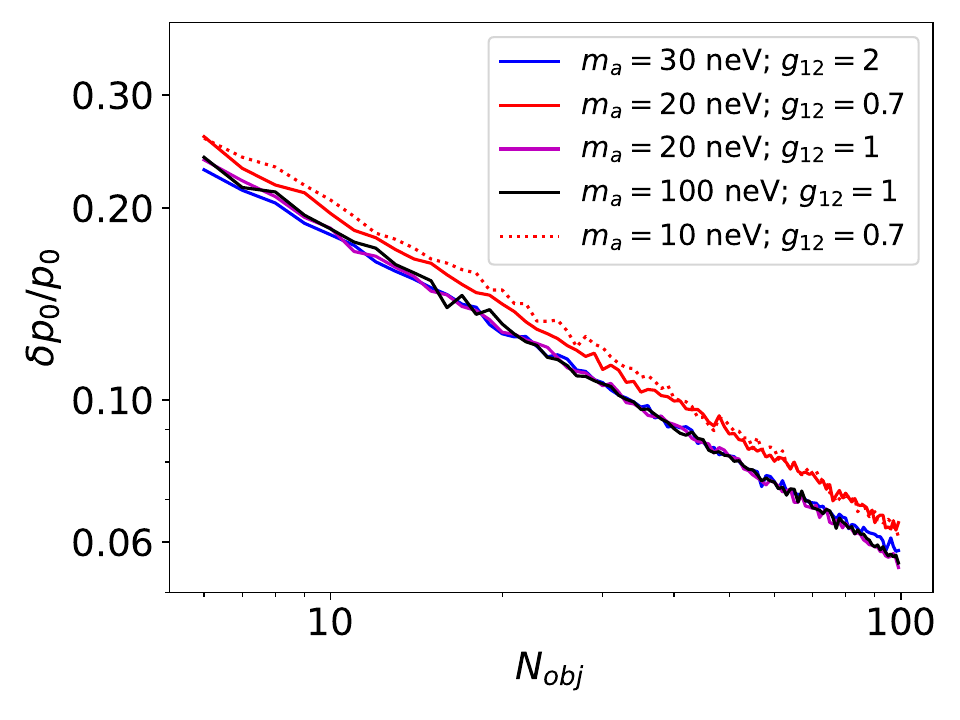}
        \caption{\textit{Left panel:} Photon survival probability $P_{\gamma\gamma} $and its averages as a function on the photon energy. Photon survival probability when passing through the galaxy cluster for different realizations of the cluster magnetic field properties (blue lines). All realizations have the same radial profile of the magnetic field but vary randomly in their orientation in the photon polarization plane and in the sizes of the magnetic domains where the field remains approximately constant. Green and red lines demonstrate the effect of averaging over 16 and 100 randomly selected realizations, respectively. Black dotted-dashed line shows the analytical approximation to these lines in case of infinite number of objects.  
        The ALP parameters for all curves are (m$_a$, g$_{a\gamma\gamma}$) = (20 neV, 7$\times$10$^{-13}$ GeV$^{-1}$). Blue curves were obtained by numerically solving ALP propagation equations via the \texttt{ALPro} code~\cite{Matthews:2022gqi}.
        \textit{Right panel:} 1$\sigma$ scatter value $\delta p_0/p_0$ as a function of the number of stacked objects for various values of the ALP mass $m_a$ and photon-ALP coupling $g_{12}\equiv g_{a\gamma\gamma} \cdot 10^{12}$~GeV. See text for further details.
        }
        \label{fig:pgg}
\end{figure}

Analytical formula for the photon-ALP conversion probability is available only for the simple case of averaging over a large number of objects with equal-size $s$ domains, constant-amplitude magnetic field which randomly changes its orientation from domain to domain.
In this case the averaged over objects photon-ALP conversion probability $\langle P_{a\gamma}\rangle$ is given by
$\langle P_{a\gamma}\rangle \simeq 1/3(1-{\rm exp}(-3 P_{a\gamma} r/2s))$~\cite{Grossman:2002by}, where $N\gg 1$ is the number of domains in each object, $r$ -- size of the object, and $P_{a\gamma}$ is a photon-ALP conversion probability in a single domain. Asymptotically as $r/s\rightarrow \infty$,  $\langle P_{a\gamma}\rangle \rightarrow 1/3$, such that the survival probability $\langle P_{\gamma\gamma}\rangle = 1- \langle P_{a\gamma}\rangle$ is greater or equal to 2/3. Generally, in this case $\langle P_{a\gamma}\rangle$ presents a step-like function of the photon energy. For $E\ll E_c$, $\langle P_{\gamma\gamma}\rangle \simeq$ 1. For $E\gg E_c$, the survival probability is always larger than 2/3 and approaches 1 with decreasing $g_{a\gamma\gamma}$, as shown by the red line in Fig.~\ref{fig:pgg} presenting the averaging over a 100 clusters with realistic magnetic field profiles. 

As mentioned in the introduction, for a single AGN-galaxy cluster pair the current methods require marginalisation over many realizations of the random magnetic field  to address the substantial uncertainties associated in the knowledge of the magnetic field. Here, we will explore the combination of a large number of sources in order to average out the details of properties of the magnetic field in galaxy clusters. 

In order to compute the photon survival probability of an AGN-emitted photon travelling through turbulent magnetic field of  a galaxy cluster, we solve Eq.~(\ref{eq:eom}) numerically using the \texttt{ALPro} code~\cite{Matthews:2022gqi}.
The cluster is discretized into domains with random transverse magnetic field orientations and coherence lengths randomly drawn between $2$ and $34 \, \text{kpc}$~\cite{Bonafede:2010xg,2015JCAP...01..011K} similarly to \cite{Malyshev:2025iis}. The transverse field in each domain is given by $B_T(r) = B(r) \cos\phi$, with $\phi$ randomly sampled to reflect angular variation and $B(r)$ corresponding to the reference magnetic field profile, see Sec.~\ref{sec:b_field}.

The survival probability has a large variance from one realization of the magnetic field to another, which results in very different oscillation patterns in the VHE range. When averaging over a large number of objects (i.e. magnetic fields realizations), the expected oscillation pattern becomes smoother and exhibits a step-like suppression making it a key feature for ALP-photon conversion in combined galaxy cluster searches. The gamma-ray spectra of AGN located behind or inside galaxy clusters may experience a  suppression in the VHE range occurring from photon-to-ALP conversion. For the set of parameters (m$_a$, g$_{a\gamma}$), we calculate $P_{a \gamma}$ and average it over a large number of realizations to obtain $\langle P_{a \gamma} \rangle$, which can be approximated~\cite{Malyshev:2025iis} as 
\begin{align}
\label{eq:pgg_model}
\langle P_{a \gamma} \rangle = p_0/(1 + (E/E_c)^k).     
\end{align}
This provides a relation between the $(p_0,E_c,k)$ parameters of the survival probability $\langle P_{\gamma\gamma}\rangle = 1 - \langle P_{a \gamma} \rangle$ and the ALP parameters $(m_a,g_{a\gamma\gamma})$.

Figure~\ref{fig:pgg} shows the effects of averaging over a different number of AGNs located behind clusters of galaxies on the photon survival probability $P_{\gamma\gamma}$ as a function of its energy for an ALP mass of $m_a$ = 2$\times$10$^{-8}$ eV and an axion-photon coupling $g_{a\gamma\gamma}$ = 7$\times$ 10$^{-13}$ GeV$^{-1}$ assuming a reference magnetic field described in Sec.~\ref{sec:b_field}. 
For a single GC, the prediction of the  expected survival pattern exhibits a strong variance due to the random projection in the transversal plane of the photon propagation direction.
Blue curves present the photon survival probabilities for 5 individual randomly selected objects.
When many AGN-GC pairs are observed, the averaged probability over the different observations becomes more and more predictable.
Green and red curves present the effect of averaging over 16 and 100 AGN, respectively.
For this number of AGN-GC pairs the scatter of $\langle P_{\gamma\gamma}\rangle$ at 1 TeV significantly reduces. 
Black dotted line illustrates the adapted model for photon survival probability $\langle P_{\gamma\gamma} \rangle$,
see Eq.~\eqref{eq:pgg_model}. 

At the same time the limited statistics of the number of objects result in a certain scatter of green curves around the parametrized model given in Eq.~\eqref{eq:pgg_model}. To account for an uncertainty connected to this scatter we additionally allowed $p_0$ in Eq.~\eqref{eq:pgg_model} to vary within the interval with a width $\delta p_0/p_0$ dependent on the number of considered objects. We define this interval in a way such that at any energy above $2\,E_c$ it contains $68\%$ of all simulated $P_{a \gamma}$. We found that $\delta p_0 /p_0$ is only weakly dependent on the ALP parameters, as shown in the right panel of Fig.~\ref{fig:pgg}. In what follows we explicitly treat $p_0$ as a free parameter that can vary around its mean value predicted by a model with $\delta p_0/ p_0$ treated as $1\sigma$ dispersion of the corresponding distribution.
The green shaded region in the right panel of Fig.~\ref{fig:pgg} stands for a $2\sigma$ containment band of the green curves. For 16 and 100 AGN pairs $1\sigma$ band can be estimated as $\sim 16$\% and $\sim 6$\%, respectively.

\subsection{Expected observed VHE gamma-ray flux}
In addition to possible photon-ALP conversion, as VHE gamma rays travel cosmological distances, their interaction with 
lower-energy photons become relevant. The low-energy photon field comes from all the light in the universe accumulated due to star formation processes and by AGNs through accretion since the epoch of reionization. It covers a wide wavelength range from the far-UV to the millimeter, and is usually referred as the Extragalactic Background Light (EBL)~\cite{Franceschini:2008tp,Dominguez:2010bv,2012MNRAS.422.3189G,Stecker:2016fsg}. 

The interaction of VHE gamma rays with EBL photons produce an exponential attenuation of the energy-differential source flux reaching the Earth via the pair production process $\gamma_{\rm VHE} + \gamma_{\rm EBL} \rightarrow e^+ + e^-$.
Therefore VHE gamma-ray spectra of extragalactic sources such as AGN show high-energy exponential reduction $\propto e^{-\tau(E, \epsilon, z)}$, where $\tau$ is the optical depth of VHE gamma rays which depends on the gamma-ray energy $E$, the EBL photon energy $\epsilon$, and the redshift of the source $z$. The precise shape depends on the spectral energy distribution of EBL photons whose determination is complex (see, \textit{e.g.}, Refs.~\cite{2013ApJ...768..197I,2014AAS...22440101D}). 

\begin{figure}[!ht]
    \centering
    \includegraphics[width=0.8\linewidth]{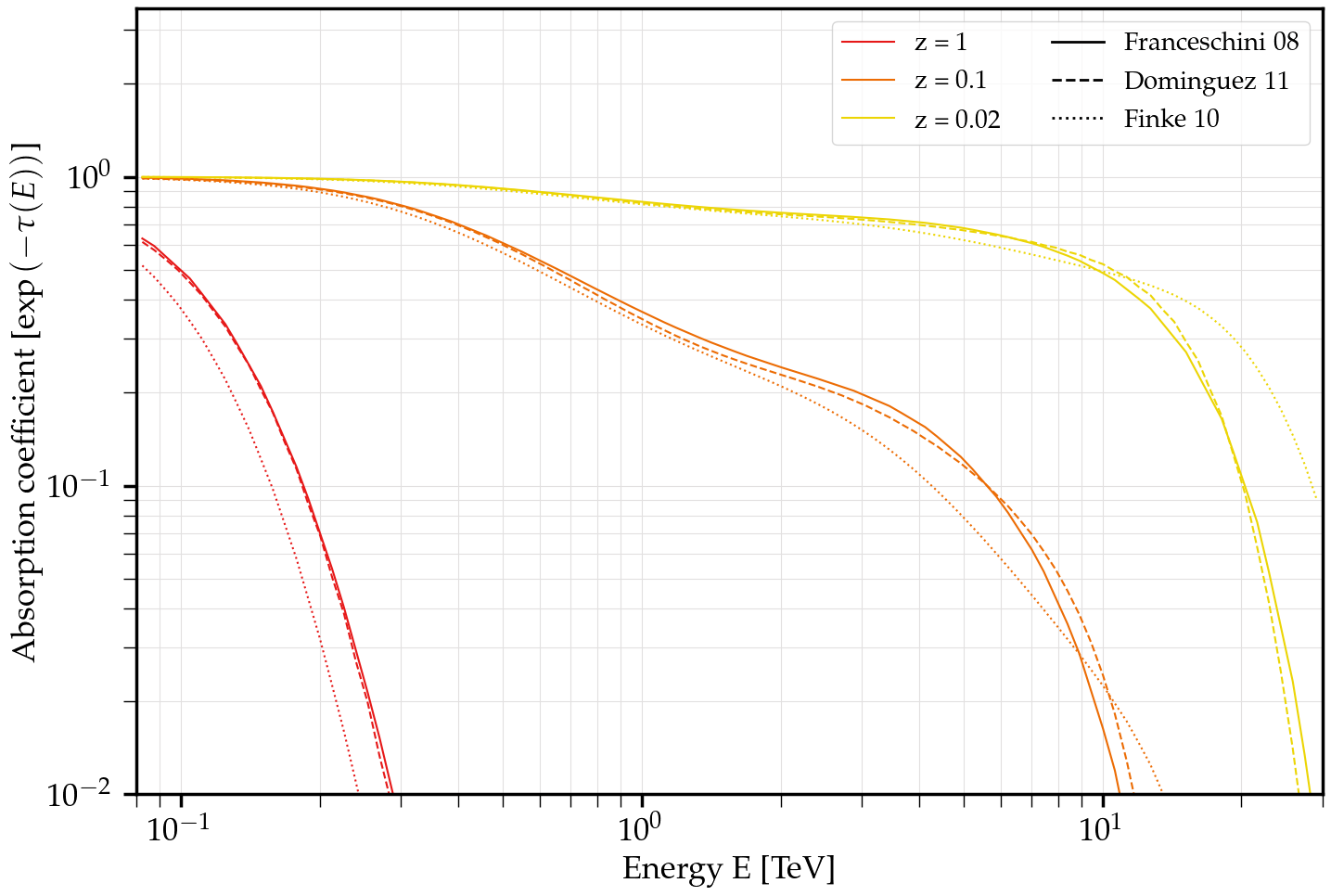}
    \caption{Comparison of the absorption coefficient $e^{-\tau(E)}$ as a function of energy $E$ for three different EBL models, Franceschini~\cite{Franceschini:2008tp} (solid line), Dominguez~\cite{Dominguez:2010bv} (dashed line) and Finke~\cite{Finke:2009xi} (dotted line), respectively. The absorption coefficient is computed for the source redshifts $z$ of 0.02, 0.5 and 1, respectively.}
    \label{fig:EBL_model_comparison}
\end{figure}

We note that EBL absorption effects can partially mimic the effects of photon–ALP conversion in stacked datasets of objects if the characteristic energy of the photon–ALP conversion feature is located close to the EBL attenuation energy range.
In what follows, we will make use of the Francheschini EBL model~\cite{Franceschini:2008tp} as a baseline model for the computation of the opacity $\tau$, while other EBL models  such as the Dominguez and Finke models~\cite{Dominguez:2010bv,Finke:2009xi}, will be used to bracket the systematic uncertainty of the IACT sensitivity to ALP searches related to this choice.
Figure~\ref{fig:EBL_model_comparison} shows the absorption coefficient $\rm exp(-\tau(E))$ as a function of gamma-ray energy $E$ for 
the three above-mentioned EBL models and source redshift $z$ values of 0.02, 0.1 and 1, respectively.

The energy-differential gamma-ray flux expected to be observed on Earth from an AGN with a GC along the line of sight can be expressed as
\begin{equation}
\frac{d\Phi^{\rm obs}_\gamma (E)}{dE}= \frac{d\Phi^{\rm int}_\gamma (E)}{dE} \times P_{\gamma\gamma}(E, m_a, B_T) \times e^{-\tau(E,\epsilon,z)}\, ,
    \label{eq:flux}  
\end{equation}
where $P_{\gamma\gamma}$ is the survival probability of the VHE gamma ray traversing the GC.
$P_{\gamma\gamma}$ = 1 in case of absence of ALP-photon conversion in the energy range of interest.

\section{Data analysis of combined AGN observations with IACTs}
\label{sec:analysis}

\subsection{Data modelling}
The model-expected number of events $N^\mathcal{M}_{ik}$ for the $k^{th}$ pair in the $i^{th}$ energy bin of width $\Delta E_i$ with a potential presence of photon-ALP conversion can be expressed as :
\begin{equation}
\label{eq:count}
N^\mathcal{M}_{ik} (m_a, g_{a\gamma\gamma}, \mathcal{P}_k) = T_{{\rm obs}, k} \int\limits_{E_i - \Delta E_i /2}^{E_i + \Delta E_i /2} dE  \int\limits_{-\infty}^{\infty} dE'\, \frac{d\Phi^\mathcal{M}_{k}(E',\mathcal{P}_k)}{dE'}\, 
    A_{{\rm eff}, k}^{\gamma}(E')\, G(E - E')\, 
\end{equation}
with the expected energy-differential flux given by
\begin{equation}
\label{eq:expectedflux}    
\frac{d\Phi^\mathcal{M}_{k}(E',\mathcal{P}_k)}{dE'} = \Phi_{0,k}\left(\frac{E'}{1\,\rm{TeV}}\right)^{-\Gamma_k}\times e^{-\tau(E',\epsilon,z)}\times P_{\gamma\gamma}(E', m_a, g_{a\gamma\gamma})\, .
\end{equation}

The finite energy resolution of the instrument $G$ is modeled as a Gaussian with width given by $\sigma/E$, where $E$ and $E’$ are the true and reconstructed energies, respectively. 
$d\Phi^M_{k}/dE'$ is the predicted flux from the AGN-cluster pair $k$ assumed to be a power law parameterised by the set of parameters $\mathcal{P}_k = (\Phi_{0,k}; \Gamma_k)$ -- with the normalisation $\Phi_{0,k}$ and spectral index $\Gamma_k$, see Sec.~\ref{sec:stat_method} for more details. 
The spectrum can possibly be affected by photon-ALP conversion $P_{a\gamma}$. In absence of ALP-photon conversion,
$P_{\gamma\gamma}(E', m_a, g_{a\gamma\gamma}=0 ) \equiv 1$.
 $T_{\rm obs,k}$ stands for the observation time of the pair $k$, and $A_{\rm eff, k}^\gamma$ corresponds to the energy-dependent effective area to VHE gamma rays for a given object $k$ due the zenith-angle dependence of $A_{{\rm eff}, k}^\gamma$. 
When no ALP-photon conversion is assumed, the energy-differential spectrum is simply extrapolated in the TeV energy range following a power-law parametrisation derived by \lat attenuated by the absorption coefficient $e^{-\tau}$. The extrapolation is made from 100 GeV to 60 TeV using 10 logarithmically-spaced spectral bins for all 41 datasets discussed in Sec.~\ref{sec:obs}. The number of expected signal in absence of ALP-photon conversion is obtained from Eq.~(\ref{eq:count}) with $P_{\gamma\gamma}$ = 1.

\subsection{Statistical analysis method}
\label{sec:stat_method}
In order to assess the sensitivity of current
IACTs to ALP coupling to photon, we construct mock datasets that realistically build upon the most recent instrumental response functions (IRF) of IACTs as discussed in Sec.~\ref{sec:obs}. We note, that for the stacked dataset analysis, the energy resolution is not playing a critical role contrary to the single target searches, where the energy resolution is of prime importance due to the rapid and strong oscillations of the $P_{\gamma\gamma}$ with energy. However, as explained in Sec.~\ref{sec:models}, when combining over many targets, the searched pattern becomes smoother such that it can be well parametrized by a step-like function.

The statistical analysis and the computation of the expected sensitivity for an IACT are performed following the definition of a log-likelihood ratio test statistic (TS). The TS exploits the expected spectral feature of the searched signal. 
The likelihood function for a given object $k$ in the $i^{\rm th}$ energy bin is expressed as $\mathcal{L}_{\rm ik}(N^\mathcal{M}_{ik},N^\mathcal{O}_{ik}) = Pois(N^\mathcal{M}_{ik},N^\mathcal{O}_{ik})$,
where $N^\mathcal{M}_{ik}$ and $N^\mathcal{O}_{ik}$ are the expected numbers of counts from the model and from the observation, respectively.  
The likelihood function for a given object $k$ is
therefore defined as $\mathcal{L}_{\rm k} = \prod_i \mathcal{L}_{\rm ik}$. 
For a combination of N observational datasets, the combined likelihood expresses as $\mathcal{L}_{\rm comb} = \prod_{k=1}^{\rm N} \mathcal{L}_{\rm k}$. 

In order to compare the models with and without ALP conversion we construct the log-likelihood ratio written as :
\begin{equation}
    \label{eq:LL}
    -2\,{\rm log}\,\lambda(m_a, g_{a\gamma\gamma}, \mathcal{P}^1,\mathcal{P}^2 ) = - 2 \log \frac{\prod\limits_{i,k} \mathcal{L}_{ik}\left(N^\mathcal{M}_{ik}(m_a, g_{a\gamma\gamma},\mathcal{P}^{1}_k), N^\mathcal{O}_{ik}\right)}   
    {\prod\limits_{i,k} \mathcal{L}_{ik}\left(N^\mathcal{M}_{ik}(m_a, g_{a\gamma\gamma}=0,\mathcal{P}^{2}_k), N^\mathcal{O}_{ik}\right)} \, .
\end{equation}
The result of the minimisation of this quantity over $\mathcal{P}^1$ -- parameters of the model with ALP, and $\mathcal{P}^2$ -- parameters of the model without ALP, corresponds to the TS of the ALP detection:
\begin{equation}
    {\rm TS}\,(m_a, g_{a\gamma\gamma}) = \min\limits_{\mathcal{P}^1, \mathcal{P}^2} -2\,{\rm log}\,\lambda(m_a, g_{a\gamma\gamma}, \mathcal{P}^1,\mathcal{P}^2 ) \, .
\end{equation}

For Asimov datasets, often used to estimate the sensitivity of the instruments, $N^\mathcal{O}_{ik} = N^\mathcal{M}_{ik}(m_a, g_{a\gamma\gamma}=0,\mathcal{P}^{LAT} )$ where $\mathcal{P}^{LAT} = (\Phi_k^{LAT}, \Gamma_k^{LAT})$ are the normalisation and spectral slope of the source $k$ derived from \lat data, see also Eq.~(\ref{eq:count}). In this case the denominator in Eq.~(\ref{eq:LL}) is optimised trivially (as optimal parameters $\Phi_k$, $\Gamma_k$ obviously correspond to the ones from \lat GeV data). Consequently during the optimisation one needs to optimise the nominator such as:
\begin{align}
& \text{TS}(m_a, g_{a\gamma\gamma}) 
    = - 2\min\limits_{\mathcal{P}^{1}} \sum\limits_{ik} \left(N^\mathcal{O}_{ik}-N^\mathcal{M}_{ik}(m_a,g_{a\gamma\gamma},\mathcal{P}^{1})\right)N^\mathcal{O}_{ik}\log\left(\frac{N^\mathcal{M}_{ik}(m_a,g_{a\gamma\gamma},\mathcal{P}^{1})}{N^\mathcal{O}_{ik}}\right)
    \end{align}
Assuming that the spectral indices remain the same between GeV and TeV bands, this TS follows a $\chi^2$ distribution with a single degree of freedom in the limit of large statistics. 
In this limit, one-sided 95\% upper limits on $g_{a\gamma\gamma}$ can be set by solving for the ALP-photon coupling above the best fit where TS = 2.71. We use this procedure in what follows to compute our sensitivity expressed as 95\% C.L. mean expected upper limit given a mock dataset.
\begin{figure}[!hb]
    \centering
    \includegraphics[width=0.8\linewidth]{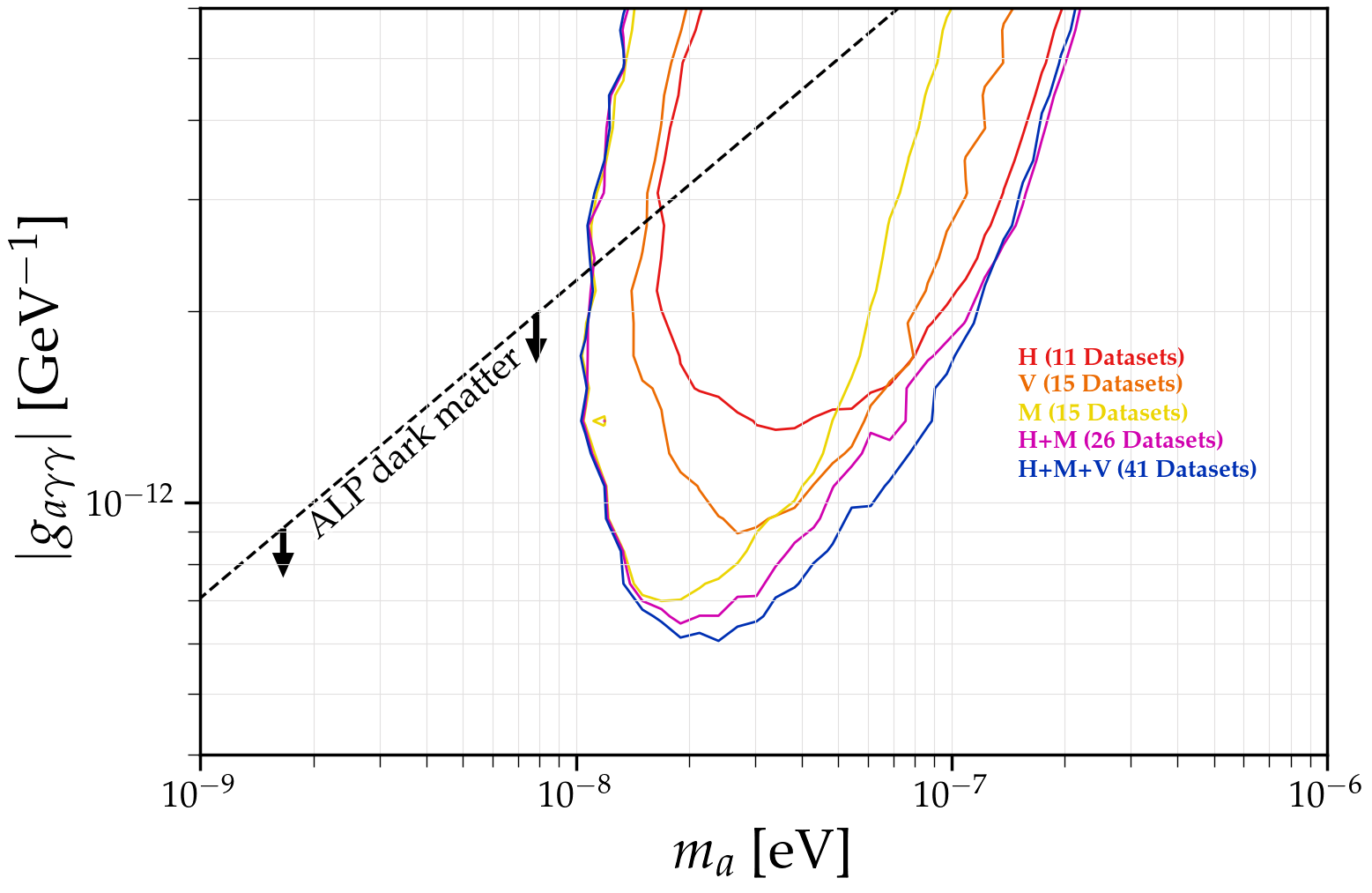}
    \caption{Sensitivity on the $g_{a\gamma\gamma}$ coupling versus ALP mass $m_a$  for combined mock datasets of IACT observations assuming the Francheschini EBL model. The sensitivity is expressed as the 95\% C. L. mean expected upper limit. The sensitivity is given for mock observations of 11 AGNs by H.E.S.S. (H, red line), 15 by VERITAS (V, orange line), 15 AGNs by MAGIC (M, yellow line). Combined mock observations by H.E.S.S. and  MAGIC (H+V, 26 datasets, pink line), and by H.E.S.S., MAGIC and VERITAS (H+M+V, 41 datasets, blue line), respectively, are also displayed. Below the dashed thick line corresponds to the region of the parameter space where ALP can comprise all the DM in the universe~\cite{Arias:2012az,Ringwald:2012hr}.}
    \label{fig:res_combineddatasets}
\end{figure}

\section{Results}
\label{sec:results}
Following the formalism introduced in the previous sections, we compute the sensitivity of a combined analysis of mock observations by the currently operating IACTs -- \hess, MAGIC and VERITAS. We start by showing how much our forecast sensitivity improves with the increased number of stacked datasets. After this, we explore the impact of systematic uncertainties from the EBL modelling on the foreseen sensitivity reach, and we assess the ability to reconstruct an injected fake ALP signal in the mock data. Then, the sensitivity reach is discussed in light of the existing constraints ranging from searches for axions from the Sun and star clusters, X-ray and gamma-ray satellites, to haloscope experiments.

Figure~\ref{fig:res_combineddatasets} shows the improvement of the expected sensitivity reach with an increased number of combined datasets, from a combined analysis of several datasets by a single IACT, \textit{e.g.}, 11 by H.E.S.S. or 15 by MAGIC, towards a combination of 41 datasets obtained by mock observations of H.E.S.S., MAGIC and VERITAS. For an ALP mass of m$_a$ = 3$\times$10$^{-8}$ eV, the sensitivity improves by a factor of about 3, reaching a ALP-photon coupling value of g$_{a\gamma\gamma}$ = 6$\times$10$^{-13}$ GeV$^{-1}$.
\begin{figure}[!ht]
    \centering
    \includegraphics[width=0.8\linewidth]{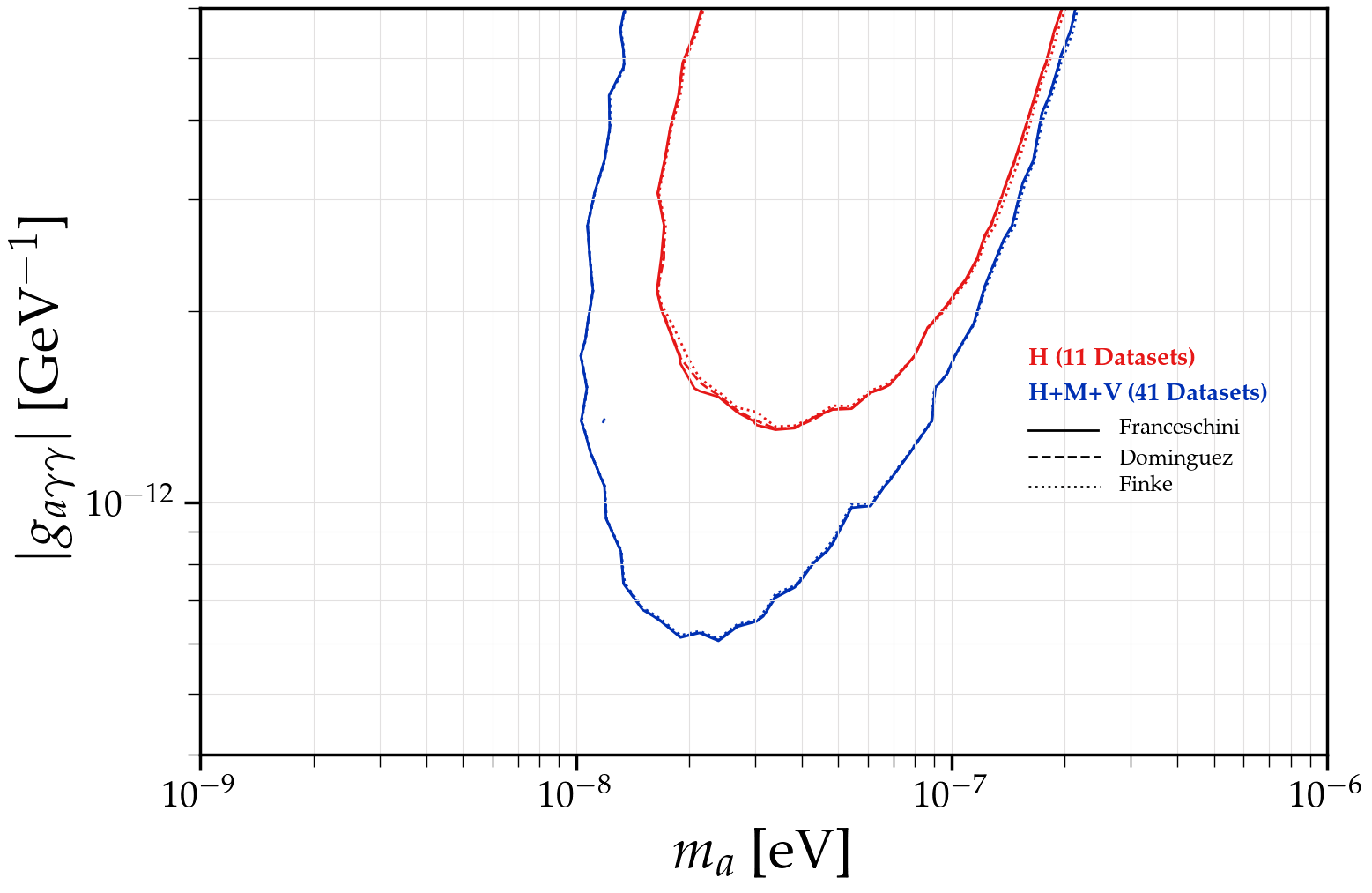}
    \caption{Sensitivity on the $g_{a\gamma\gamma}$ coupling versus ALP mass $m_a$  for combined mock datasets of IACT observations for different EBL models, \textit{i.e.}, the Francheschini (solid line), Dominguez (dashed line), and Finke (dotted line) models, respectively. The sensitivity is expressed as the 95\% C. L. mean expected upper limit. The sensitivity is given for mock observations by H.E.S.S. of 11 AGNs (reds line) and combined mock observations by H.E.S.S., MAGIC and VERITAS of 41 AGNs (blue line), respectively.}
    \label{fig:res_EBLimpact}
\end{figure}
A sharp decrease in sensitivity with ALP mass around 10$^{-8}$ eV is observed as the critical energy gets closer to the IACTs low-energy threshold. For masses of a few 10$^{-7}$ eV, the sensitivity weakens as the critical energy falls into the strong EBL absorption regime and consequent poor statistics of the data.
Such an approach enables us to probe the region of the parameter space  where ALPs can account for all the cold dark matter in the Universe via the vacuum-realignment mechanism\textbf{~\cite{Arias:2012az,Ringwald:2012hr}}.
\begin{figure}
        \centering
        \includegraphics[keepaspectratio, width=0.48\linewidth, clip]{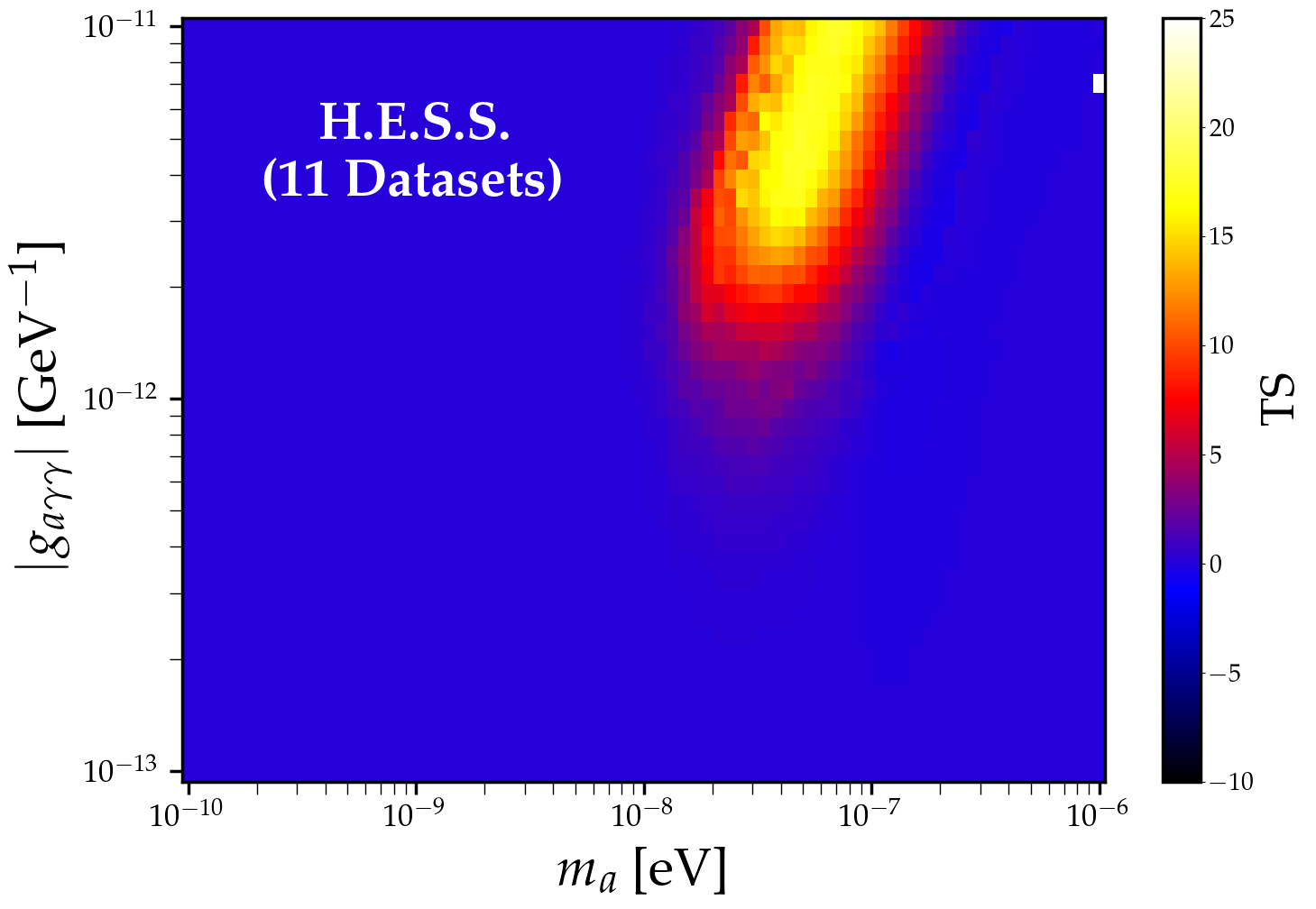}
        \includegraphics[keepaspectratio, width=0.48\linewidth, clip]{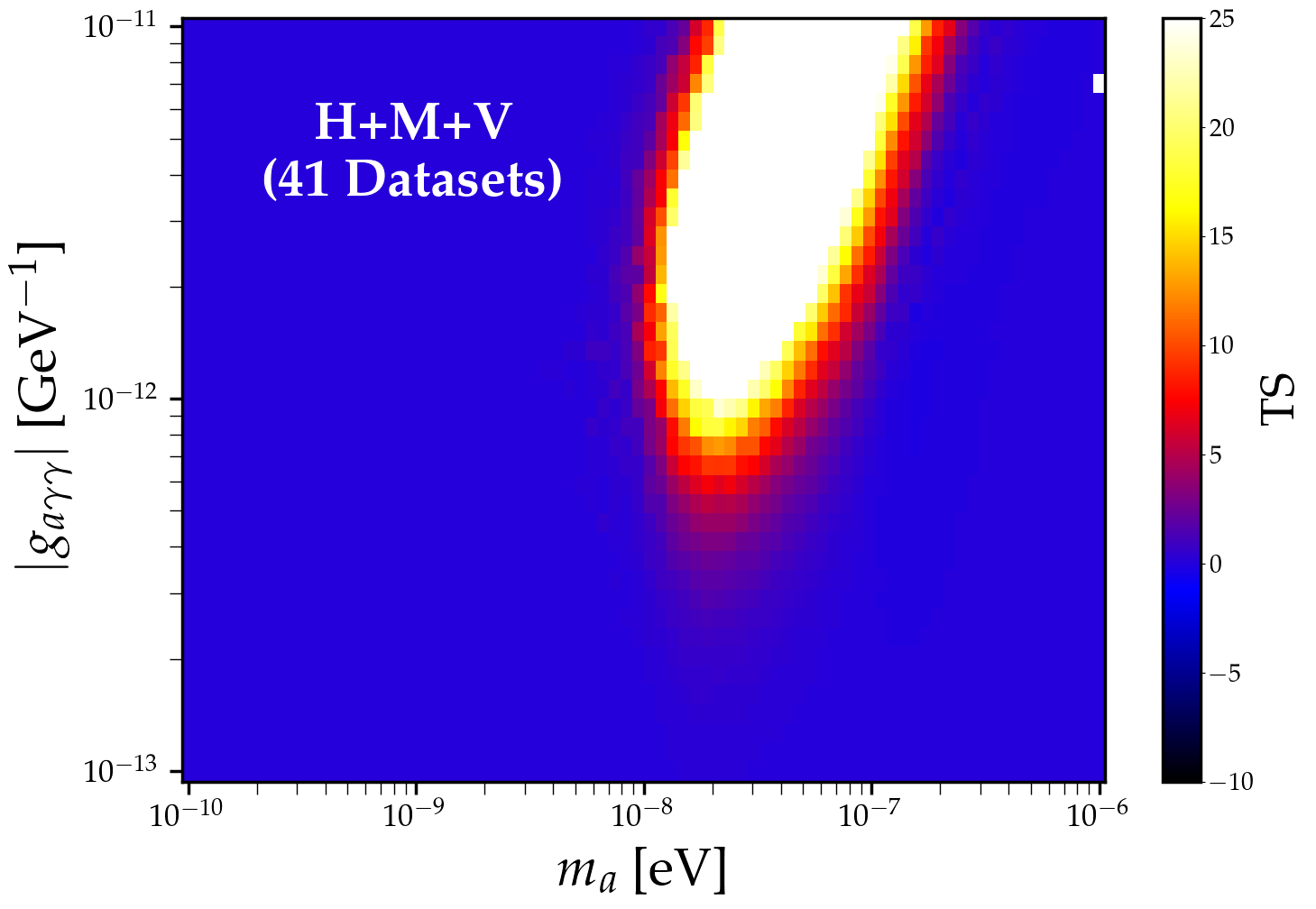}
        \includegraphics[keepaspectratio, width=0.48\linewidth, clip]{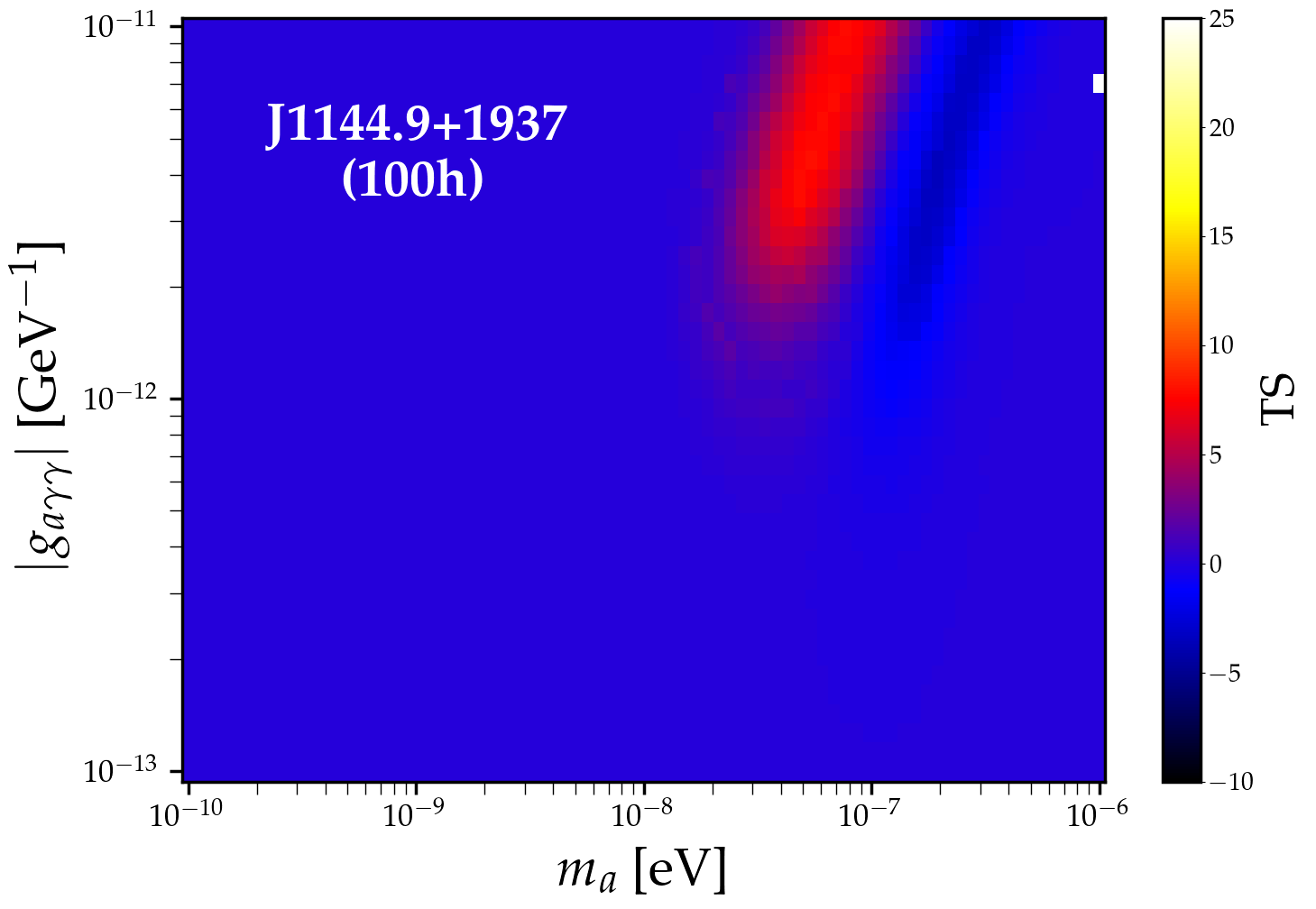}
        \includegraphics[keepaspectratio, width=0.48\linewidth, clip]{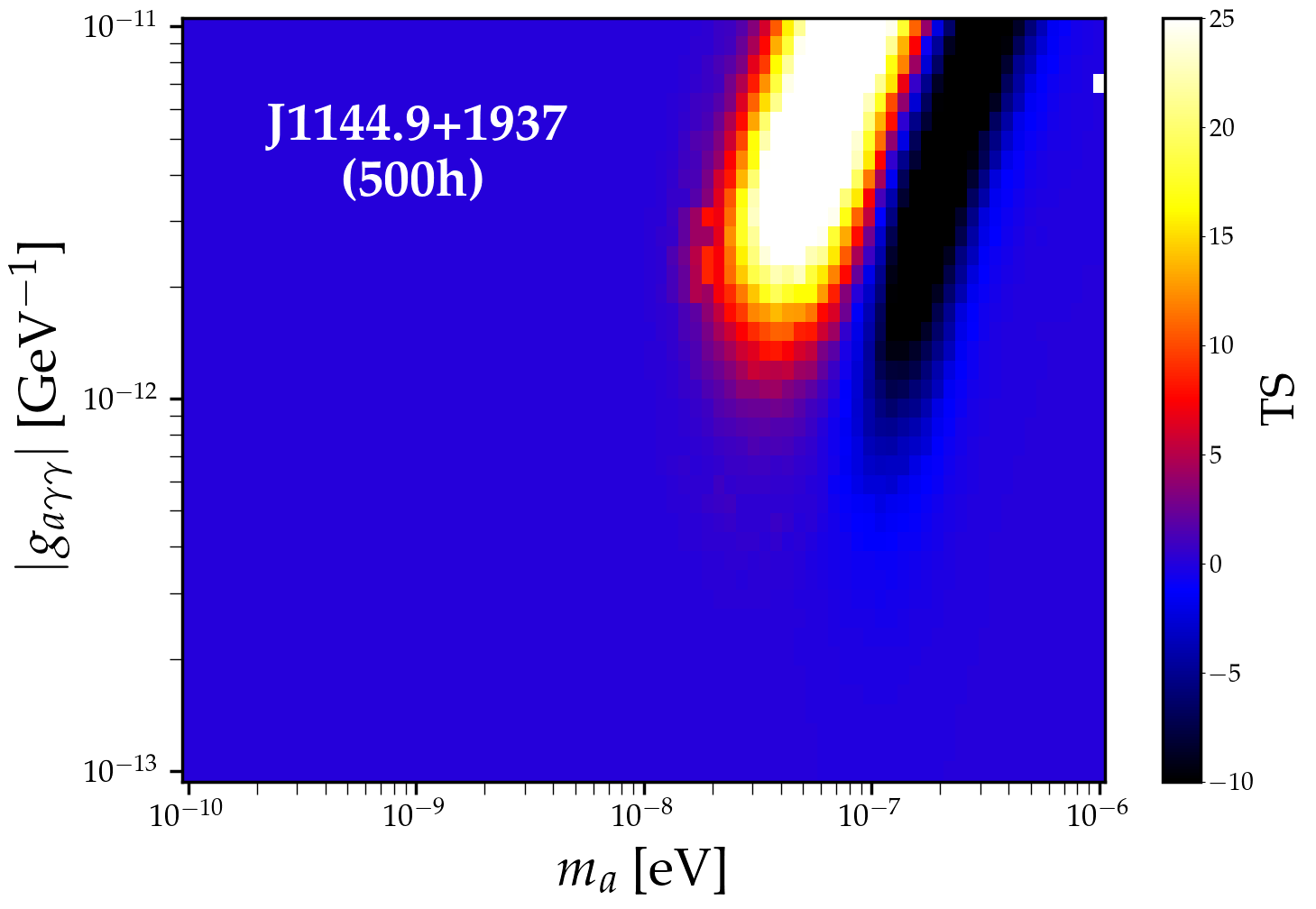}
        \caption{\textit{Top left panel}: TS map expressed in the ($g_{a\gamma\gamma}$, $m_a$) for mock observations of 11 datasets seen by H.E.S.S. for an exposure of 50h for each. The mock observations are simulated with the Francheschini EBL model, and the computation of the sensitivity to $g_{a\gamma\gamma}$ as a function of $m_a$ assumes the Finke EBL model. 
        \textit{Top right panel}: TS map for mock observations of 41 datasets with combined observation by H.E.S.S., MAGIC and VERITAS.
        \textit{Bottom panels}: TS map for mock observations of the J1144.9+1937 (z=0.02) AGN by H.E.S.S. for an observation time of 100 h on the left and 500 h on the right.
        }
        \label{fig:res_EBLmismodelling}
\end{figure}

To investigate the possible bias connected to the exact choice of EBL model we consider several up-to-date EBL absorption models -- Dominguez \textit{et al.}~\cite{Dominguez:2010bv}, Franceschini \textit{et al.}~\cite{Franceschini:2017iwq}, and Finke \textit{et al.}~\cite{Finke:2009xi}. 
Figure~\ref{fig:res_EBLimpact} shows the impact of the choice of the EBL model on the expected sensitivity for several considered datasets and all discussed EBL models. 
The impact of the EBL model is limited to 3\% below a mass of 7$\times$10$^{-8}$ eV, and increases to 6\% for masses up to 2$\times$10$^{-7}$ eV considering a stacking analysis of 41 datasets.

To mimic the effect of EBL mismodelling we have simulated the mock data with the \textit{Francheschini} EBL model and perform the computation of the sensitivity by maximizing the likelihood assuming the \textit{Finke} EBL model.
The top panels of Fig.~\ref{fig:res_EBLmismodelling} show the TS values in the $(m_a,g_{a\gamma\gamma})$ plane for 11 (\hess only) and  41 (\hess + MAGIC + VERITAS) datasets, respectively. A weak detection (with TS $\sim -0.50$) appears in the case of the analysis of 11 datasets.
This false-detection of the signal is connected to the different energy-dependencies of the EBL attenuation in Finke and Francheskini models.
Below 10~TeV energies, the Finke EBL model induces more gamma-ray attenuation for the sources at redshift $z\lesssim$ 0.2 compared to the Francheschini model. Such sources at relatively low redshifts drive the overall sensitivity. Therefore, an increased number of gamma rays is required at energies of $\lesssim$ 10 TeV, which can be obtained with an ALP-photon coupling of a few 10$^{-12}$ GeV$^{-1}$ and a ALP mass of about 10$^{-7}$~eV as the step-like decrease of the survival probability will occur at energies above $\sim$10 TeV. We note a presence of a negative-TS value that mimics the appearance of an ALP signal. 
For 41 (\hess+ MAGIC + VERITAS) datasets the redshifts of the considered sources extend to the higher values which reduces the relative impact of low-redshift sources and consequently the effect of EBL mismodelling.

We further investigate the case of the dataset dominated by a large number of hours of observations performed on a single AGN. 
The bottom panels of Fig.~\ref{fig:res_EBLmismodelling} show the TS values in the $(m_a,g_{a\gamma\gamma})$ plane, assuming that the dataset is dominated either by 100 or 500 hours of exposure, respectively, on the source J1144.9+1937 at $z=0.02$. As the photon statistics increases, the false detection becomes more and more significant. This shows that the strong observing time bias towards a single bright and close-by object  is prone to make the EBL mismodelling impact more and more pronounced.
We argue thus that the observational program to be performed shall result in statistically homogeneous dataset, without a strong preference to individual bright/nearby objects. In this case the effects of the potential EBL mismodelling are minimized and the program provides a promising observational strategy.

\begin{figure}
    \centering
\includegraphics[width=0.8\linewidth]{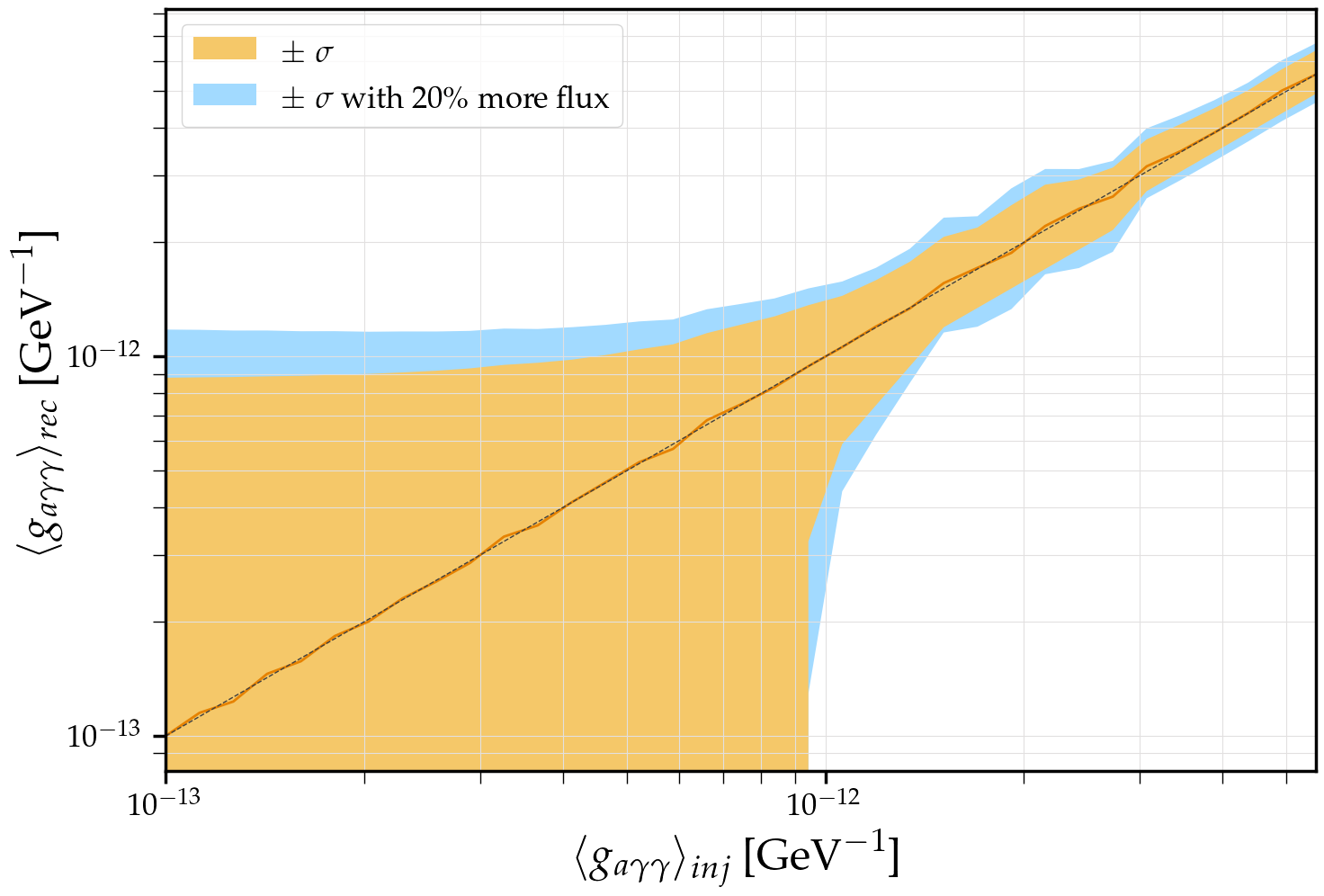}
    \caption{Reconstruction of an injected ALP signal in the 41 datasets from mock observation by H.E.S.S., MAGIC and VERITAS. For an ALP mass of 7$\times$10$^{-8}$ eV we inject a specific ALP-photon coupling $g_{a\gamma\gamma}^{\rm inj}$, and consider our ability to reconstruct it, labeled as $g_{a\gamma\gamma}^{\rm reco}$. The orange line and band show the mean and 1$\sigma$ reconstruction, respectively, in the absence of systematic uncertainty. Adding a systematic uncertainty on the energy-differential flux normalisation of 20\% expands the 1$\sigma $ uncertainty band as shown by the blue-shaded area. 
    }
    \label{fig:res_performance}
\end{figure}

In order to explore our ability to reconstruct an ALP signal present in the data, we inject an ALP-photon conversion signal for the coupling value $g_{a\gamma\gamma}^{\rm inj}$ in each of the considered objects. We attempt then to reconstruct the coupling
$g_{a\gamma\gamma}^{\rm reco}$ by maximizing the likelihood following the procedure described in Sec.~\ref{sec:stat_method}. The results of this analysis are shown in Fig.~\ref{fig:res_performance}. Firstly, we show our ability to reconstruct the signal in the absence of systematic uncertainties. For an ALP mass of 7$\times$10$^{-8}$ eV, the coupling values above $10^{-12}$~GeV$^{-1}$ can be recovered with a bias better than 4\% in the considered coupling range. For lower values of couplings, only upper limits can be derived.
It highlights that the statistical uncertainty on the recovery of the ALP-photon coupling  across the values we show is considerable.
In order to further explore the reconstruction power, we inject a known ALP signal into the mock data and explore our ability to reconstruct it in the presence of 20\% systematic uncertainties on the flux normalisation.
For an injected ALP-photon coupling value $g_{a\gamma\gamma}^{\rm inj}$ = 2$\times$10$^{-12}$ GeV$^{-1}$, the 1$\sigma$ band increases by 14\% compared to the case with statistical uncertainty only.

\begin{figure}
    \centering
    \includegraphics[width=\linewidth]{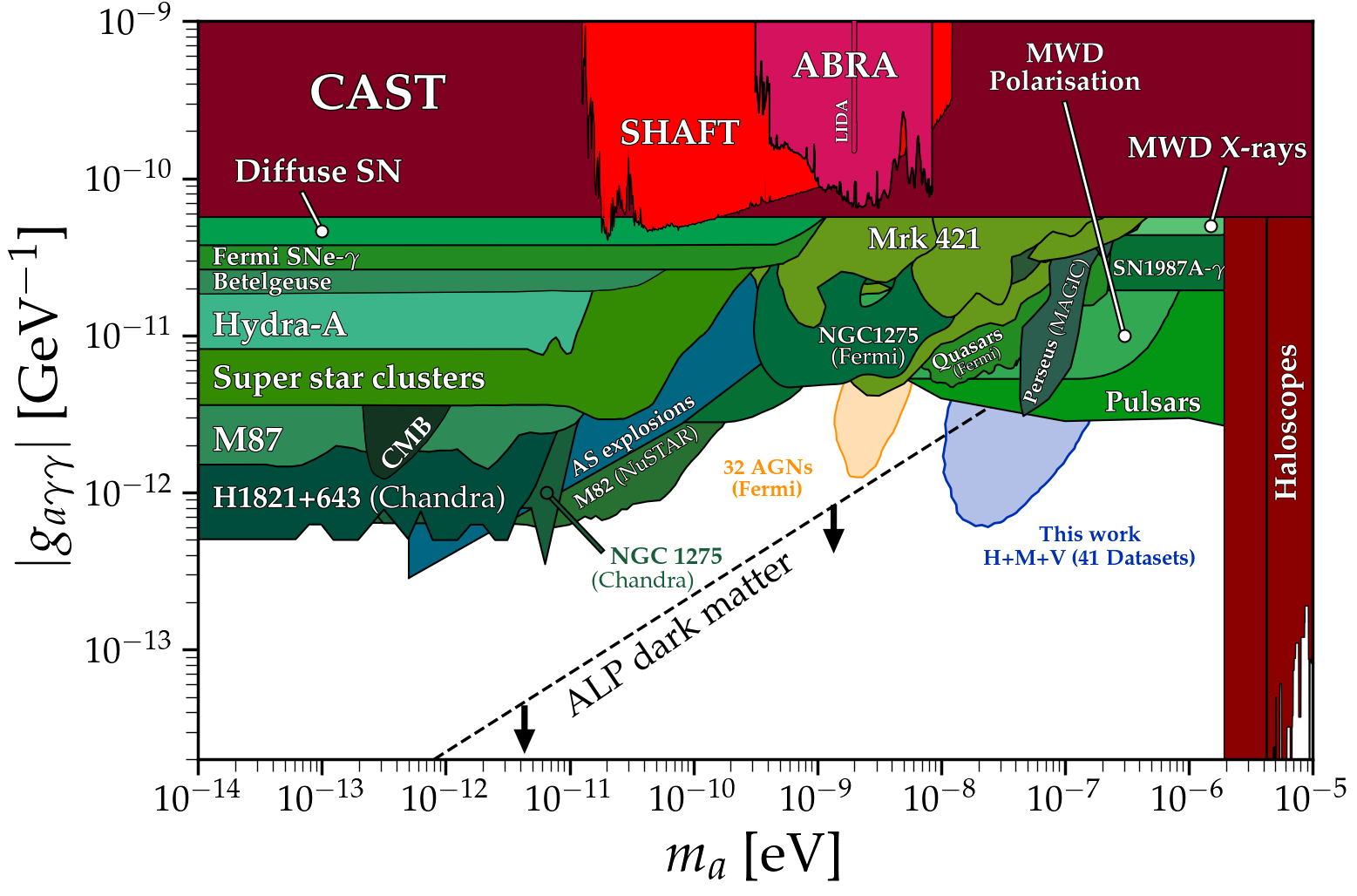}
    \caption{
    Sensitivity on the ALP-photon coupling $g_{a\gamma\gamma}$ versus the ALP mass $m_a$. The sensitivity is expressed as 95\% C.L. mean expected upper limits. The sensitivity is  derived from combined mock observations by H.E.S.S., MAGIC and VERITAS, resulting in 41 independent datasets (solid blue line).  Also shown are the  exclusion regions from current experiments extracted from Ref.~\cite{axionlimits}. The region below the dashed thick line indicates parameter values for which ALPs can account for the observed cold dark matter abundance via the vacuum misalignment mechanism, varying the initial misalignment angle value and allowing a general variation of ALP mass with temperature~\cite{Arias:2012az,Ringwald:2012hr}.   
    95\% C.L. upper limits derived from the stacked analysis of observations of 32 AGN-GC pairs with \lat and studied in
Ref.~\cite{Malyshev:2025iis} are shown as the solid orange line.
    }
    \label{fig:summaryplot}
\end{figure}

Finally, Figure~\ref{fig:summaryplot} shows the expected sensitivity reached by combined mock observations by H.E.S.S., MAGIC and VERITAS in light of the present constraints from a wealth of approaches, from dedicated experiments such as CAST and SHAFT to astrophysical searches in X rays (e.g., Chandra) and gamma rays (e.g, \lat, and IACTs), radio observations of pulsars, and haloscopes, extracted for Ref.~\cite{axionlimits} and references therein. Our approach enables to  probe yet uncovered region of the parameter space where ALP can comprise all the DM, 
with the strongest sensitivity reached at $g_{a\gamma\gamma}$ = 5$\times$10$^{-13}$ GeV$^{-1}$ for an ALP mass of $m_a = 2\times$10$^{-8}$\,eV.
The region of the parameter space where ALPs can
account for the observed cold dark matter abundance via the vacuum misalignment
mechanism, assuming an initial misalignment angle $\theta_i = \mathcal{O}(1)$ and standard cosmology is given by the thick dashed line. Below this line ALPs can constitute up to a 100\% of cold dark matter varying the initial misalignment angle and allowing a variation of the ALP mass with temperature~\cite{Arias:2012az,Ringwald:2012hr}.

\section{Conclusions and outlook}
\label{sec:conclusions}
In this work, we have performed detailed studies of the 
sensitivity of the current generation of IACTs, \textit{i.e.}, \hess, MAGIC and VERITAS, to ALP searches deploying a new analysis strategy utilizing a combination of mock observations of AGNs located behind galaxy clusters. We performed a selection of AGNs from \lat catalogue with galaxy clusters and suitable for further VHE observations.

With a state-of-the-art analysis framework making use of a log-likelihood ratio test statistics, we assess the sensitivity to the ALP-photon coupling via a combined analysis. While the survival gamma-ray probability pattern versus energy is almost unpredictable for a single object, averaging over a large number of objects makes the oscillation pattern smoother exhibiting a step-like suppression in the VHE regime. With 41 datasets of mock AGN observations, we have demonstrated that with such an approach we can probe yet uncharted ALP parameter space in the 10$^{-8}$ - 10$^{-7}$ eV mass range and coupling $g_{a\gamma\gamma}\gtrsim 5\times 10^{-13}$~GeV$^{-1}$. This range is of a great potential interest, as ALPs in this mass-couplings range can constitute up to a 100\% of the dark matter in the Universe.

In drawing these conclusions, we have assumed 50 hours of observation per datasets. This requires an intensive observational program of 550 hours for H.E.S.S., and 750 hours for MAGIC and VERITAS observatories, respectively. Given the relatively large required total observational time for a single IACT we foresee that the observations can be performed within observational campaigns split over a few to several years.

The analysis of the limiting uncertainties to the ALP DM reach shows that even for relatively long observational time of $50$~h per target, the statistical uncertainty remains dominant,
demonstrating the importance of the continued data collection with ongoing IACTs. We further explore the systematic uncertainty effects induced by EBL modelling and systematics on source flux measurements. 
While we found that the impact of EBL model choice is negligible we found that EBL mismodelling could induce false-detection of the ALP signal. High photon statistics obtained from 
statistically-homogenous observations of selected AGNs distributed over a large range of redshifts tends to significantly reduce such an effect.

We note also several other sources of systematic uncertainties potentially relevant for the derived sensitivities on ALP parameters. These include possible biases, \textit{e.g.}, in the modelling of the magnetic field in the clusters~\cite{Malyshev:2025iis} or in AGN flux-spectral slope dependency~\cite{Malyshev:2025iis}. The uncertainties connected to the systematics of these types could typically reach a factor of two in the sensitivity to the coupling constant $g_{a\gamma\gamma}$~\cite{Malyshev:2025iis}, but can be mitigated with the future studies of the cluster-averaged magnetic field profiles either with numerical N-body clusters-simulations~\cite[\textit{e.g.},][]{eagle,tng,tng-cl} or, 
\textit{e.g.}, future observations with forthcoming radio instruments such as SKA~\cite{ska}. The possible bias connected to the mismodelling of the spectral slope can be neglected if the VHE observations are accompanied by the quasi-simultaneous observations at other wavelengths (\textit{e.g.}, GeV and/or X-rays) which will allow the multiwavelength modelling and firm reconstruction of the spectral slope. At the same time we foresee that the number of AGN-cluster pairs can potentially increase in the future due to either discovery of a higher number of VHE AGNs (\textit{e.g.}, due to a discovery of hard-spectrum and/or flaring objects) or due to the detection of a higher number of galaxy clusters with present-day (\textit{e.g.}, eROSITA), or future missions. Such an increase will allow either even better constraints in ALP parameter space or eventual discovery of ALP.

Our results demonstrate that the proposed stacking analysis strategy of the AGN-GC pairs provides a viable strategy that can probe previously uncharted and physically-motivated region of ALP DM parameter space even with the present-day IACTs. The results derived in this work further motivate near-future observations of AGN-GC pairs with the forthcoming CTA observatory in the quest of ALP DM.

\acknowledgments
The authors acknowledge support by the state of Baden-W\"urttemberg through bwHPC.

\bibliographystyle{JHEP}
\bibliography{bib.bib}

\end{document}